\documentclass[runningheads]{llncs}
\usepackage[T1]{fontenc}
\usepackage{comment}
\usepackage[numbers]{natbib}
\usepackage{pdflscape,tabularx,booktabs,enumitem}
\usepackage{longtable} 
\usepackage{multirow}
\usepackage{graphicx}
\usepackage{epstopdf}
\usepackage{float}
\usepackage{xcolor}
\usepackage{amssymb}
\usepackage{array}

\newlist{tightitemize}{itemize}{1}
\setlist[tightitemize]{nosep, topsep=0pt, leftmargin=*, label=\textbullet}
\usepackage{tabulary}
\usepackage{array}
\usepackage{ragged2e}
\newcolumntype{P}[1]{>{\RaggedRight\arraybackslash}p{#1}}

\renewcommand{\arraystretch}{0.9}
\usepackage{url}

\begin{document}

\title{AI in Education Beyond Learning Outcomes: Cognition, Agency, Emotion, and Ethics}
\titlerunning{AI in Education Beyond Learning Outcomes}

\author{Lucile Favero\inst{1}\orcidID{0000-1111-2222-3333} \and \\
Juan Antonio Pérez-Ortiz\inst{2}\orcidID{1111-2222-3333-4444} \and \\
Tanja Käser\inst{3}\orcidID{2222--3333-4444-5555}
\and \\
Nuria Oliver\inst{1}\orcidID{0000-1111-2222-3333}}
\authorrunning{Favero L. et al.}

\institute{ELLIS Alicante, Spain \and
Universitat d'Alacant, Spain \and
École Polytechnique Fédérale de Lausanne, Switzerland \email{lucile@ellisalicante.org}}

\maketitle

\begin{abstract}
Artificial intelligence (AI) is rapidly being integrated into educational contexts, promising personalized support and increased efficiency. However, growing evidence suggests that the uncritical adoption of AI may produce unintended harms that extend beyond individual learning outcomes to affect broader societal goals. This paper examines the societal implications of AI in education through an integrative framework with four interrelated dimensions: cognition, agency, emotional well-being, and ethics. Drawing on research from education, cognitive science, psychology, and ethics, we synthesize existing evidence to show how AI-driven cognitive offloading, diminished learner agency, emotional disengagement, and surveillance-oriented practices can mutually reinforce one another. We argue that these dynamics risk undermining critical thinking, intellectual autonomy, emotional resilience, and trust, capacities that are foundational both for effective learning and also for democratic participation and informed civic engagement. Moreover, AI’s impact is contingent on design and governance: pedagogically aligned, ethically grounded, and human-centered AI systems can scaffold effortful reasoning, support learner agency, and preserve meaningful social interaction. By integrating fragmented strands of prior research into a unified framework, this paper advances the discourse on responsible AI in education and offers actionable implications for educators, designers, and institutions. Ultimately, the paper contends that the central challenge is not whether AI should be used in education, but how it can be designed and governed to support learning while safeguarding the social and civic purposes of education.

\end{abstract}

\section{Introduction}

\begin{quote}
\emph{“\textbf{Sapere aude}! Have courage to use your own reason!”} \\
\hfill --- Immanuel Kant, \textit{An Answer to the Question: What is Enlightenment?} (1784)
\end{quote}

Artificial Intelligence (AI) is increasingly recognized as a transformative tool for education. According to UNESCO, AI has the potential to tackle some of the most pressing educational challenges, enhance teaching and learning practices, and accelerate progress toward the Sustainable Development Goal 4 (SDG 4) \cite{unescoAIeducation}, which aims to ``ensure inclusive and equitable quality education and promote lifelong learning opportunities for all" by 2030. Yet, the rapid development and unprecedented adoption rates of AI---and particularly chatbots---in education present significant societal risks that often outpace existing policy and regulatory frameworks.

\begin{figure}[ht]
\centering
\includegraphics[width=0.5\textwidth]{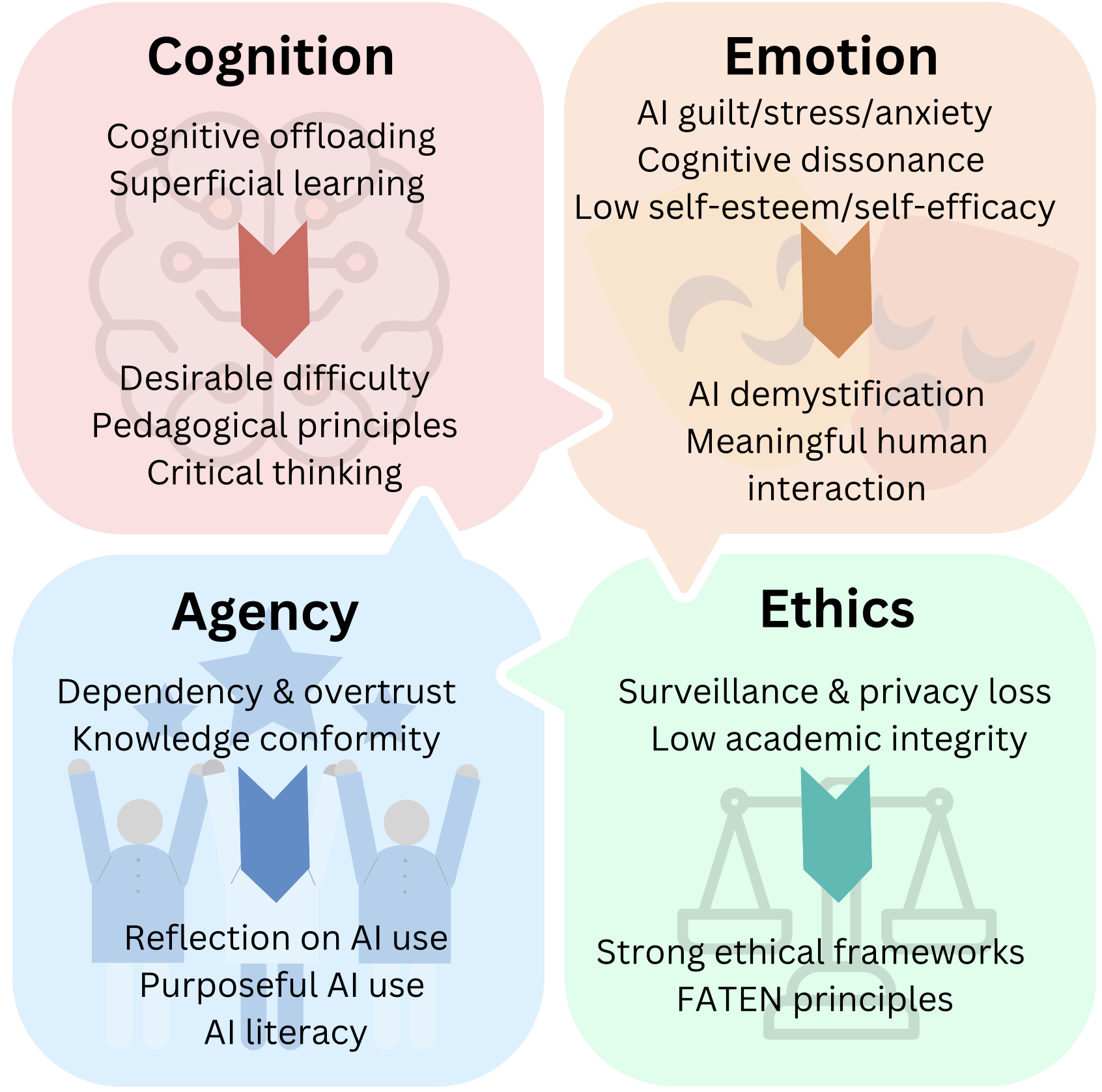}
\caption{Ensuring a human-centered, responsible, and critical use of AI in education requires the integration of four interrelated dimensions: cognition, agency, emotion and ethics.}
\label{fig:visual}
\end{figure}

Empirical evidence supports the educational potential benefits of AI-driven systems. Intelligent Tutoring Systems (ITS), for instance, have shown generally positive effects on student learning and performance \cite{letourneauSystematicReviewAIdriven2025}. Generative AI could also serve as a personal tutor---particularly valuable in contexts of teacher shortages---and produce adaptive, personalized, diverse, and up-to-date educational materials tailored to individual learners’ needs \cite{al-zahraniExploringImpactArtificial2024}. These systems aim to promote self-paced, autonomous learning by supporting the key phases of self-regulated learning: goal setting, performance monitoring, and reflection \cite{vorobyevaPersonalizedLearningAI2025,al-zahraniExploringImpactArtificial2024}. Instant feedback and 24/7 availability should support the learner’s ability to progress independently and at their own pace \cite{vorobyevaPersonalizedLearningAI2025}. Consequently, AI has the potential to reduce teacher workload, increase cost-effectiveness, and personalize and scale educational access, contributing to a more equitable and democratized learning environment \cite{worldbank2024ai}. 

Despite its many advantages, the integration of AI in education raises profound societal concerns: while AI tools are rapidly being integrated into educational settings, the public and scholarly discourse on their unintended harms and social implications remains fragmented. Previous research often focuses on discrete issues---such as ethical risks \cite{williams2024ethical}, cognitive effects \cite{gerlichAIToolsSociety2025a}, or surveillance practices \cite{kwapiszPrivacyConcernsStudent2024}---without integrating them into a cohesive framework that considers the broader societal consequences. In response, this paper provides a multidisciplinary analysis of the use of AI in education from a societal perspective, examining its impacts on cognition, emotion, agency and ethics. Beyond identifying potential harms, it proposes actionable strategies for educators, institutions and policymakers to ensure that AI integration supports learning while safeguarding the students' well-being, autonomy and fundamental rights.

\section{Unintended Harms of AI in Education}\label{sec:challenges}

We identify and analyze four interrelated dimensions where AI integration can inadvertently undermine educational outcomes from a student perspective: cognition, agency, emotional well-being, and ethics, depicted in Figure~\ref{fig:visual}. 

\subsection{Cognition} \label{sec:co}

Effective learning depends on \emph{active engagement} and \emph{cognitive effort}, which are essential for fostering independent reasoning, critical thinking, and long-term intellectual resilience. Cognitive offloading, \emph{i.e.}, the use of external tools to reduce mental load, can support learning when it frees resources for deeper reflection. However, over-reliance on AI for analytical, reasoning, or synthesis tasks risks undermining independent thought, memory, creativity, and motivation \cite{ravseljHigherEducationStudents2025}. Large-scale studies show that high dependence on AI correlates with lower performance on critical thinking assessments, mediated by cognitive offloading \cite{gerlichAIToolsSociety2025a}. These effects threaten not only educational outcomes but also the civic reasoning skills necessary for informed democratic participation, making learners more vulnerable to misinformation, echo chambers, and political polarization \cite{pennycook2021psychology,lazer2018science}.

AI’s educational impact is closely tied to whether it embodies pedagogical principles, such as active learning, constructivism, and scaffolding. \emph{Active learning} engages students in analysis, synthesis, and evaluation \cite{freeman2014active}, \emph{constructivism} emphasizes knowledge construction through experience and context \cite{fosnot2013constructivism}, and \emph{scaffolding} provides temporary support that cultivates autonomy and independent problem-solving \cite{van2002scaffolding}. Many widely used AI tools, such as chatbots, provide pre-digested answers in a passive, directive manner, limiting opportunities for reflection and critical engagement \cite{letourneauSystematicReviewAIdriven2025, ravseljHigherEducationStudents2025}. In contrast, AI tutoring systems explicitly designed according to these principles (managing cognitive load, scaffolding reasoning, and promoting meta-cognition) can dramatically enhance learning outcomes \cite{kestinAITutoringOutperforms2024, zerkouk2025comprehensive}. The distinction is thus between pedagogically aligned AI vs. naive AI, with the former capable of nurturing autonomy, critical thinking, and societal readiness.

A complementary cognitive element is \emph{effortful learning} and desirable difficulties. Cognitive struggle, \emph{i.e.}, mental effort that may feel uncomfortable, is essential for deep understanding, long-term retention, and transferable knowledge \cite{bjork2011making}. Students often misinterpret effort as a sign of poor learning, a bias known as the \emph{illusion of fluency}, which drives preference for low-effort, superficially satisfying activities \cite{zhangYouHaveAI2024}. AI tools that prioritize speed, fluency, and convenience---such as open-domain chatbots---risk amplifying this bias, reducing cognitive struggle and weakening critical thinking \cite{zhaiEffectsOverrelianceAI2024}. To counteract this, AI must preserve and strategically manage cognitive friction, prompting retrieval, offering delayed feedback, and framing uncertainty as an opportunity for inquiry. When AI scaffolds productive struggle rather than bypassing it, it fosters intellectual resilience, independent reasoning, and the critical thinking skills vital for both education and democratic society.

\subsection{Agency}\label{sec:agency}
In education, \emph{agency} refers to a learner´s ability to make intentional, informed and autonomous choices, underpinned by self-regulation, meta-cognition and critical thinking \cite{roe2024generative}. As AI tools become pervasive in academic contexts, their convenience and persuasive outputs can erode agency, such that students become passive recipients of algorithmically generated content \cite{al-zahraniUnveilingShadowsHype2024}. This risk is amplified when AI outputs are inconsistent, biased, or false, potentially reinforcing inequalities and spreading misinformation, with implications for both educational equity and societal discourse \cite{becirovic2025exploring,zhangYouHaveAI2024,roe2024generative}. 

An important unintended harm is \emph{dependency}, in which students rely excessively on AI, undermining independent problem-solving, decision-making and critical evaluation \cite{lanQualitativeSystematicReview2025}. Over time, this dependency reduces the learners' ability to assess, question and engage with information autonomously, while impairing meta-cognition and intellectual autonomy \cite{gerlichAIToolsSociety2025a}. Vulnerable populations, such as students with lower digital literacy, are particularly at risk, as AI systems can act as opaque authorities, increasing susceptibility to misinformation and educational disparities \cite{becirovic2025exploring}. 

Dependency often leads to \emph{overtrust}. The apparent reliability and human-like presentation of AI tools encourage students to accept its outputs uncritically \cite{gerlichPowerVirtualInfluencers2023,schaaff2024impacts}. This cognitive surrender reduces evaluative thought, deep learning and confidence in personal decision-making \cite{williams2024ethical}. Explicit instruction about the statistical nature of today´s AI algorithms and their limitations can restore critical engagement, supporting independent reasoning and evaluative autonomy \cite{bastaniGenerativeAICan2024}.

A perhaps subtler, long-term threat is that of \emph{intellectual conformity}. By providing ready-made answers, AI tools can subtly nudge learners toward normative reasoning, discouraging creativity and diverse interpretations \cite{tafazoliExploringPotentialGenerative2024}. Over time, this conformity risks standardizing thought patterns instead of fostering imaginative reasoning. 

Agency is also a societal necessity. Critical thinking, autonomy and evaluative judgment cultivated in educational environments are at the core of our capacity to discern misinformation, to identify and resist manipulation and to participate meaningfully in democratic societies \cite{pennycook2021psychology}. Therefore, AI tools that undermine the students' agency also threaten these broader civic competencies. Conversely, AI tools that support reflection, critical evaluation and diverse choices can strengthen both individual and societal resilience. 

\subsection{Emotion}\label{sec:emo}

Prolonged or uncritical reliance on AI can have negative emotional impact. \emph{Technostress} arises when learners must adapt to opaque, rapidly evolving AI systems without sufficient control or understanding, leading to anxiety and cognitive overload \cite{tarafdar2007impact}. \emph{Digital fatigue} and disengagement emerge when AI mediates most learning interactions, reducing opportunities for meaningful social connection. In self-paced or AI-dominated environments, students may experience diminished belonging and emotional resilience—outcomes that disproportionately affect already vulnerable learners and risk amplifying educational inequities \cite{klimova2025exploring}.

AI use also interacts strongly with \emph{self-efficacy} and \emph{self-esteem}. Students with lower confidence are more likely to rely on AI as a compensatory strategy, creating a self-reinforcing cycle: avoidance of challenge reduces confidence, which in turn increases dependence on AI \cite{lanQualitativeSystematicReview2025, rodriguez-ruizArtificialIntelligenceUse2025}. When learners perceive AI as inherently superior than them, motivation, creativity, and independent thinking may erode, fostering impostor syndrome and long-term disengagement \cite{chanExploringFactorsAI2024}. Conversely, students who understand AI’s limitations report higher self-efficacy, highlighting the protective role of AI literacy \cite{becirovic2025exploring}.

A further emotional tension arises in the form of \emph{AI guilt} and \emph{cognitive dissonance}. Many students experience discomfort, shame, or anxiety when AI use conflicts with values of authenticity, effort, and academic integrity \cite{chanExploringFactorsAI2024}. This unresolved tension, feeling both assisted and inauthentic, can undermine the learners' sense of identity and belonging in academic communities \cite{lanQualitativeSystematicReview2025}. Conversely, some students exhibit \emph{AI entitlement}, viewing algorithmic assistance as a rightful expectation rather than a pedagogical choice \cite{ravseljHigherEducationStudents2025}. The coexistence of guilt-driven anxiety and entitlement-driven normalization reflects a broader normative breakdown around effort, authorship, and fairness in AI-mediated education.

From a societal perspective, these emotional dynamics matter. Education plays a central role in shaping both skilled workers, and confident, resilient, and socially grounded citizens. Emotional disengagement, diminished self-belief, and fractured norms around responsibility and effort risk weaken the psychological foundations necessary for civic participation, collaboration, and trust in democracy. Mitigating these harms requires placing emotional and social well-being at the core of AI integration. Rather than replacing human connection, AI should scaffold it, supporting collaboration, reflective feedback, and dialogue. Institutions should promote AI literacy that demystifies system capabilities, normalize transparent and reflective AI use, and establish shared norms that balance innovation with authenticity. When designed and governed responsibly, AI can support the learners’ confidence, belonging, and emotional resilience, strengthening not only education, but the social fabric it sustains.

\subsection{Ethics}\label{sec:ethic}

The integration of AI, especially chatbots, into educational environments raises profound ethical challenges, particularly concerning student privacy, surveillance, academic integrity and the dynamics of power in digital learning spaces. 
These issues demand urgent attention as institutions increasingly adopting generative AI tools need to consider the pedagogical, legal, and psychological implications. 
AI-driven educational tools often rely on continuous data collection, creating \emph{privacy} challenges and environments where students may feel constantly monitored, evaluated, and recorded \cite{al-zahraniUnveilingShadowsHype2024, shores2024surveillance}. Such surveillance has psychological and pedagogical consequences: when learners perceive that every interaction is stored or analyzed, they may avoid experimentation, intellectual risk-taking, and learning through error, processes essential for critical thinking and deep understanding \cite{mason2016learning, meraUnravelingBenefitsExperiencing2022}. Fear of being wrong can lead to shame, reduced confidence, and conformity, undermining education’s role in cultivating curiosity, resilience, and reflective judgment \cite{kwapiszPrivacyConcernsStudent2024, mezirowTransformativeLearningTheory1997}.
From a societal perspective, this erosion of safe learning spaces is particularly troubling. Education is one of the few institutions tasked with preparing individuals to question authority, challenge assumptions, and participate meaningfully in democratic life. Pervasive surveillance risks normalizing compliance over critique.

These concerns are compounded by the handling of \emph{student data}. Generative AI systems process highly sensitive information (\emph{e.g.}, academic performance, behavioral patterns, and personal interactions) often under opaque governance structures \cite{al-zahraniExploringImpactArtificial2024, klimova2025exploring}. Students typically have limited understanding of how their data is collected, stored, or monetized, creating a profound power imbalance between learners, institutions, and technology providers \cite{kwapiszPrivacyConcernsStudent2024}. Even when personal data protection regulations exist, such as GDPR\footnote{The European General Data Protection Regulation: https://eur-lex.europa.eu/eli/reg/2016/679/oj/eng} in Europe, enforcement is uneven, and consent is frequently nominal rather than informed \cite{williams2024ethical, howladerFactorsInfluencingAcceptance2025}. As a result, tools framed as educational support can function as mechanisms of large-scale data extraction and behavioral surveillance.
Ethically responsible AI in education must therefore protect students’ rights not only to privacy, but also to make mistakes without permanent records, profiling, or reputational harm.

Ethical risks also emerge when AI is poorly integrated pedagogically. In the absence of clear guidance, students often turn to generic AI tools as shortcuts, encouraging surface-level engagement, procrastination, and practices that undermine \emph{academic integrity} \cite{niloyAIChatbotsDisguised2024}. Advanced generative models blur traditional distinctions between original and assisted work, challenging conventional plagiarism detection and assessment norms \cite{williams2024ethical}. When assessments prioritize polished outputs over learning processes, AI can become an enabler of misconduct rather than intellectual growth, threatening the credibility of educational institutions and the shared norms of fairness, effort, and trust on which they depend. Ethical AI use is thus inseparable from pedagogical design and institutional governance.

Addressing these challenges requires more than technical fixes. Institutions must embed \emph{ethical principles} into AI deployment by ensuring informed consent, data minimization, transparency, and the protection of learners’ fundamental rights. Equally important is aligning AI use with pedagogy: integrating AI literacy, defining clear and context-sensitive boundaries for acceptable use, and designing assessments that reward critical engagement, reflection, and process rather than mere output \cite{al-zahraniUnveilingShadowsHype2024}. Approaches such as iterative work, peer review, and oral defenses can reposition AI as a support for learning rather than a substitute for thinking.

Ultimately, an ethical use of AI in education is about preserving education as a space for autonomy, critical inquiry, and democratic formation. If AI systems undermine these values, they risk weakening not just learning outcomes, but the social institutions that depend on educated, reflective, and empowered citizens.

\subsection{Student Perspectives}\label{sec:stu}
In support of a human-centered approach to the integration of AI in education, it is crucial to give voice to the students’ perspectives. 

\paragraph{AI Adoption.}
According to OpenAI, young adults aged 18–24 are the largest demographic group of adopters of ChatGPT in the U.S., with over one third using the tool regularly \cite{openai2025aiready}. Among these users, more than a quarter of the messages are related to education, including tutoring, writing support, and programming help. In a survey of 1,200 students in this age group, AI tools were most often used for initiating papers and projects (49\%), summarizing texts (48\%), brainstorming (45\%), topic exploration (44\%), and revising writing (44\%). On a global scale, a 2024 international survey of 4,000 university students in 16 countries found that students value AI primarily for its timely support (63\%), help in understanding tool use (46\%), and access to training opportunities (45\%) \cite{digitaleducationcouncil2025}. 

\paragraph{Concerns and Unintended Harms.}
Students are aware of the potential negative consequences of their AI use and formulate their concerns: 61\% worry about data privacy, while many question the reliability of AI-generated content and the academic risks of overreliance. They fear that excessive dependence on AI could undermine learning, critical thinking, and the instructional value of education \cite{digitaleducationcouncil2025}. Aligned with these findings, \cite{gerlichAIToolsSociety2025a} found that students were increasingly aware of: (a) their dependency on AI for both routine and cognitive tasks; (b) the diminished opportunities for independent thinking; and (c) ethical issues such as bias, transparency, and decision-making influence. 

\paragraph{The Educational Level Shapes AI Critical Use.}
Attitudes toward AI use and the ability to critically evaluate its outputs vary significantly depending on educational background \cite{gerlichAIToolsSociety2025a}. Learners with stronger prior academic preparation or digital literacy are generally better positioned to question, contextualize, and strategically use AI outputs, while others may treat AI as an authoritative source. These disparities risk reinforcing existing cognitive and educational inequalities. These contrasting perspectives highlight a pressing need for inclusive AI literacy education that equips all learners, not only the most advantaged, with the skills to critically assess AI-generated content, understand its limitations, and engage with it as a tool rather than a substitute for thinking.

\section{Discussion}\label{sec:disc}

This paper has examined the unintended harms of AI integration in education through four interrelated dimensions: cognition, agency, emotional well-being, and ethics. Our analysis suggests that the central risk of AI in education is not technological failure, but misalignment with the social, pedagogical, and civic purposes of education. When AI systems prioritize efficiency, fluency, and automation over effort, reflection, and autonomy, they risk undermining the very capacities that education seeks to cultivate in learners and, by extension, in democratic societies.

A key insight emerging from our work is that these dimensions do not operate independently. \emph{Cognitive offloading} reduces opportunities for \emph{effortful reasoning}; \emph{diminished effort} contributes to \emph{illusory learning} and \emph{overtrust}; overtrust weakens learner \emph{agency}; and reduced agency exacerbates emotional harms such as \emph{anxiety}, diminished \emph{self-efficacy}, and \emph{dependency}. These individual-level effects are further amplified by ethical and structural factors, including surveillance, opaque data practices, and unclear institutional norms. Together, they form a reinforcing cycle in which learners become less confident, less critical, and more reliant on algorithmic authority.

From a societal perspective, these dynamics are particularly concerning. Education plays a foundational role in preparing individuals to evaluate information critically, tolerate uncertainty, engage in reasoned disagreement, and participate meaningfully in civic life. Our findings suggest that widespread reliance on AI systems that bypass cognitive struggle, normalize passive consumption, or promote conformity may erode the skills required to navigate misinformation, polarization, and manipulation in algorithmically mediated public spheres. In this sense, the risks of AI in education extend beyond learning outcomes to the health of democratic institutions and public discourse.

At the same time, our analysis does not support a rejection of AI in education. Rather, it underscores that AI’s impact is contingent on design, pedagogy, and governance. A consistent pattern emerges from the four dimensions: harms arise when AI replaces human judgment, effort, and interaction, whereas benefits emerge when AI scaffolds these processes. Pedagogically aligned AI---which is designed to manage cognitive load without eliminating struggle, to prompt reflection rather than deliver answers, and to preserve human relationships---can strengthen intellectual resilience rather than weaken it. Similarly, ethical AI governance that prioritizes privacy, transparency, and the right to make mistakes can help maintain trust and psychological safety in learning environments. A concrete illustration of this approach is \texttt{Maike} \cite{favero2024enhancing}, a privacy-preserving, environmentally sensitive educational chatbot which leverages AI to guide learners through critical questioning and self-reflection by means of the Socratic method. \texttt{Maike} exemplifies how AI systems can support epistemic agency and cognitive engagement instead of undermining them. 

This work contributes to the AIED literature by moving beyond fragmented accounts of risk to offer an integrated, societally grounded framework. Prior research has often examined ethical concerns \cite{kwapiszPrivacyConcernsStudent2024}, cognitive effects \cite{gerlichAIToolsSociety2025a}, or emotional \cite{rodriguez-ruizArtificialIntelligenceUse2025} impacts in isolation. By bringing these dimensions together, we show how unintended harms are systemic rather than incidental, and how addressing them requires coordinated interventions at multiple levels: AI design, pedagogy, assessment, institutional policy, and student AI literacy. Several implications follow. For \emph{researchers}, there is a need to study AI not only as a learning tool but as a social actor that shapes norms, identities, and power relations. For \emph{educators} and \emph{designers}, the challenge is to resist the temptation of frictionless automation and instead design for productive struggle, agency, and reflection \cite{favero2024enhancing}. For \emph{institutions} and \emph{policymakers}, clear norms, transparent data practices, and participatory governance are essential to ensure that AI supports equitable education rather than undermining it.

Finally, this discussion highlights an important normative stance: education should not optimize learners for compliance with intelligent systems, but prepare them to question, understand, and responsibly use them. If AI in education succeeds only in making learning faster or easier, it risks hollowing out the deeper purposes of education. If, however, it is aligned with cognitive resilience, learner agency, emotional well-being, and ethical responsibility, AI can contribute not only to better learning, but to a more informed, reflective, and resilient society. To support this vision, we propose in Table~\ref{tab:stakeholder_checklist} a practical checklist to help  key stakeholders---namely AI designers, students and institutions---assess whether educational AI aligns with cognitive and pedagogical principles, according to the four dimensions studied in this paper.

\begin{table}[t]
\centering
\footnotesize
\caption{Stakeholder checklist to assess AI use across cognitive, pedagogical, emotional, and ethical principles. Subsections span all stakeholders.}
\label{tab:stakeholder_checklist}
\renewcommand{\arraystretch}{1.1}
\begin{tabular}{p{1cm} p{3.5cm} p{3.5cm} p{3.5cm}}
\toprule
\textbf{Dim.} & \centering \textbf{AI Designers} & \textbf{Students} & \textbf{Institutions} \\
\midrule
\textbf{Cog.} & \multicolumn{3}{c}{\textbf{Cognitive offloading \& critical thinking}} \\
\cline{2-4}
 & $\square$ Scaffold reasoning, problem-solving, and metacognition \newline
 $\square$ Preserve cognitive effort; encourage productive struggle &
 $\square$ Avoid using AI to replace analytical, reasoning, or synthesis tasks \newline
 $\square$ Engage in effortful reasoning and reflection &
 $\square$ Design assessments that reward independent critical thinking \newline
 $\square$ Create learning activities that reinforce reflection \\
\cline{2-4}
 & \multicolumn{3}{c}{\textbf{Pedagogical \& cognitive-science principles}} \\
\cline{2-4}
 & $\square$ Align AI with active learning, scaffolding, constructivist principles \newline
 $\square$ Provide adaptive feedback instead of static content &
 $\square$ Engage actively with AI; avoid passive consumption \newline
 $\square$ Use AI to deepen exploration rather than shortcut learning &
 $\square$ Integrate AI to promote active engagement and metacognition \newline
 $\square$ Encourage productive struggle \\
\cline{2-4}
 & \multicolumn{3}{c}{\textbf{Superficial learning vs. desirable difficulties}} \\
\cline{2-4}
 & $\square$ Preserve cognitive friction; do not optimize only for fluency \newline
 $\square$ Encourage resilience and deeper understanding &
 $\square$ Recognize the difference between ease of use and actual learning \newline
 $\square$ Embrace cognitive struggle as part of deep learning &
 $\square$ Frame AI as a partner in fostering resilience and long-term understanding \\
\midrule
\textbf{Agen.} & \multicolumn{3}{c}{\textbf{Dependency}} \\
\cline{2-4}
 & $\square$ Encourage purposeful use rather than blind reliance \newline
 $\square$ Support users with low digital literacy &
 $\square$ Avoid overreliance on AI; practice independent learning \newline
 $\square$ Reflect on AI outputs &
 $\square$ Teach critical AI literacy \newline
 $\square$ Preserve student autonomy by offering engagement choices \\
\cline{2-4}
 & \multicolumn{3}{c}{\textbf{Overtrust}} \\
\cline{2-4}
 & $\square$ Communicate AI limits; highlight statistical, not infallible nature \newline
 $\square$ Make biases and limitations visible &
 $\square$ Exercise independent judgment \newline
 $\square$ Be aware AI can provide biased, false, or misleading outputs &
 $\square$ Develop informed skepticism and evaluative independence \\
\cline{2-4}
 & \multicolumn{3}{c}{\textbf{Conformity}} \\
\cline{2-4}
 & $\square$ Avoid promoting homogeneous responses \newline
 $\square$ Encourage diverse perspectives &
 $\square$ Avoid copying AI tone or structure \newline
 $\square$ Seek originality and creativity beyond AI suggestions &
 $\square$ Emphasize intellectual diversity \newline
 $\square$ Encourage critical thinking and questioning of assumptions \\
\bottomrule
\end{tabular}
\end{table}

\begin{table}[t]
\centering
\footnotesize
\renewcommand{\arraystretch}{0.9}
\label{tab:stakeholder_emotion_ethics}
\begin{tabular}{p{1.1cm} p{3.5cm} p{3.5cm} p{3.5cm}}
\toprule
\textbf{Dim.} & \textbf{AI Designers} & \textbf{Students} & \textbf{Institutions} \\
\midrule
\textbf{Emot.} & \multicolumn{3}{c}{\textbf{Emotional risks}} \\
\cline{2-4}
 & $\square$ Keep interface simple and avoid unnecessary complexity \newline
 $\square$ Implement safeguards against overuse and fatigue &
 $\square$ Monitor stress or fatigue from AI use \newline
 $\square$ Balance AI use with peer/instructor interaction &
 $\square$ Complement rather than replace human learning \newline
 $\square$ Train educators to detect technostress and support well-being \\
\cline{2-4}
 & \multicolumn{3}{c}{\textbf{Self-efficacy \& self-esteem}} \\
\cline{2-4}
 & $\square$ Clearly communicate system limits and capabilities &
 $\square$ Reflect on when AI use helps or harms confidence \newline
 $\square$ Recognize the value of own effort without comparison to AI &
 $\square$ Foster independence, critical thinking, and sense of human worth in activities \\
\cline{2-4}
 & \multicolumn{3}{c}{\textbf{AI guilt \& cognitive dissonance}} \\
\cline{2-4}
 & $\square$ Encourage reflection rather than blind reliance \newline
 $\square$ Frame AI as a support tool, not a solution provider &
 $\square$ Align AI use with personal academic values \newline
 $\square$ Acknowledge AI contributions transparently &
 $\square$ Present AI as a learning tool \newline
 $\square$ Require AI use statements or reflective assignments \\
\cline{2-4}
 & \multicolumn{3}{c}{\textbf{AI entitlement \& normative tension}} \\
\cline{2-4}
 & $\square$ Highlight appropriate vs. inappropriate AI uses &
 $\square$ Respect institutional AI policies \newline
 $\square$ Reflect on fairness and authorship &
 $\square$ Establish clear, context-sensitive AI use policies \newline
 $\square$ Facilitate open dialogue on authorship, fairness, and integrity \\
\midrule
\textbf{Eth.} & \multicolumn{3}{c}{\textbf{Privacy \& surveillance}} \\
\cline{2-4}
 & $\square$ Minimize unnecessary monitoring &
 $\square$ Feel free to experiment and make mistakes \newline
 $\square$ Understand what data is collected &
 $\square$ Limit surveillance practices \newline
 $\square$ Protect students’ right to make mistakes without penalty \\
\cline{2-4}
 & \multicolumn{3}{c}{\textbf{Data exploitation}} \\
\cline{2-4}
 & $\square$ Store sensitive data securely \newline
 $\square$ Align collection with governance and legal compliance &
 $\square$ Know who can access your data and how it is used \newline
 $\square$ Feel in control of personal information &
 $\square$ Enforce strong governance and GDPR compliance \\
\cline{2-4}
 & \multicolumn{3}{c}{\textbf{Academic integrity \& pedagogical design}} \\
\cline{2-4}
 & $\square$ Discourage shortcuts and promote genuine learning \newline
 $\square$ Design outputs to support critical thinking rather than plagiarism &
 $\square$ Respect academic integrity when using AI &
 $\square$ Embed AI literacy and guidelines in the curriculum \newline
 $\square$ Redesign assessments to reward authentic learning (peer review, oral defense, iterative writing) \newline
 $\square$ Foster an ethical academic culture through dialogue \\
\bottomrule
\end{tabular}
\end{table}

\section{Conclusion} \label{sec:conclusion}
This paper has examined the unintended harms of AI integration in education through four inter-related dimensions: cognition, agency, emotion and ethics. We argue that misaligned AI use can undermine both the learning outcomes and the social and civic aims of education, particularly when systems favor efficiency, fluency, and automation over effort, autonomy, and reflection, eroding critical thinking, learner agency, emotional resilience, and trust. Conversely, when grounded in pedagogical theory, ethical governance, and human-centered design, AI can scaffold intellectual growth while preserving the capacities essential for democratic participation. A key contribution of this work is its integrative perspective. Rather than treating cognitive, emotional, ethical, and agency-related concerns in isolation, we show how these dimensions interact to produce systemic risks that require coordinated responses across AI design, pedagogy, assessment, and institutional policy. This framing highlights that the question is not whether AI should be used in education, but how it should be designed and governed to support learning, equity, and societal resilience. 


\section*{Acknowledgements}
This work has been partially supported by a nominal grant received at the ELLIS Unit Alicante Foundation from the Regional Government of Valencia in Spain (Resolución of the Generalitat Valenciana, Conselleria de Innovación, Industria, Comercio y Turismo, Dirección General de Innovación). L.F. has also been partially funded by the Bank Sabadell Foundation. 

\section*{Declaration of Interest Statement}
The authors declare that they have no known competing financial interests or personal relationships that could have appeared to influence the work reported in this paper.

\end{document}